\documentclass[a4paper]{jpconf}
\usepackage{graphicx}

\begin{document}
\title{Possibility to enhance teraherz emission from intrinsic Josephson junction by external local heating}

\author{Hidehiro Asai and Shiro Kawabata}

\address{Electronics and Photonics Research Institute, National Institute of Advanced Industrial Science and Technology (AIST), Tsukuba, Ibaraki 305-8568, Japan}

\ead{hd-asai@aist.go.jp}

\begin{abstract}
We theoretically propose a practical method for realizing intense terahertz (THz) emission from intrinsic Josephson junctions (IJJs) using an external heat source. 
An artificial inhomogeneous temperature distribution by the local heating strongly excites the Josephson plasma wave inside IJJs and enhances THz emission power.
We  show optimum heating conditions for achieving high power THz emission. 
Our result indicates that local heat control is a powerful method to realize practical solid-state THz-emitters utilizing IJJs.
\end{abstract}

\section{Introduction}

 
%

%
\ \  Recently,  cuprate high-$T_{\rm c}$ superconductors have attracted much attention as promising candidates for compact solid-state terahertz (THz) sources.  
The layered structure in cuprate superconductor behaves as a natural stack of Josephson junctions referred to as intrinsic Josephson junctions (IJJs), and THz wave from cuprate superconductor is induced by the AC Josephson effect in IJJs. 
As a mechanism that induces THz emission, the coherent flow of Josephson-vortex was proposed first in Ref \cite{IJJKoyamaTachiki} in the presence of an applied magnetic field. On the other hand, THz emission can be also realized by the excitation of electromagnetic cavity modes in the IJJs without an applied magnetic field \cite{kinkHu1,inphaseKoshelev,fdtdKoyama1,reviewSavelev,inphaseAsai}.  After an experimental realization of intense THz emission from $\textrm{Bi}_2\textrm{Sr}_2\textrm{CaCu}_2\textrm{O}_{8+\delta}$ (Bi2212) single crystal without an applied magnetic field\cite{Ozyuzer}, THz emission from mesa-structured IJJ  in zero magnetic field condition has been intensively studied\cite{Ozyuzer,jpsjKado,hot2Wang,TBenseman}.
%
%
However, the emission powers of these IJJ emitters are much lower than 1 mW which is required for practical applications.
Thus, further investigation toward the high power emission is indispensable to realize IJJ THz emitter for practical use.

 Recently, a $hot$ $spot$ in the mesa where the temperature is locally higher than the superconducting critical temperature $T_c$   has been observed during THz emission processes\cite{hot2Wang,TBenseman}. 
 Hence, the temperature inhomogeneity such as the hot spot is considered to play a crucial role in the strong THz emission. 
Since the critical current density $j_c$ depend on the temperature,  the appearance of hot spots indicates the inhomogeneous  $j_c$ distribution.
Previous theoretical studies show such an inhomogeneous $j_c$ distribution strongly excites the Josephson plasma wave inside IJJs\cite{inphaseKoshelev,inphaseAsai}.
Therefore, the artificial control of the local temperature by some external heat source is expected to be a promising method to realize intense THz emission.
In this study, the THz emission from an IJJ mesa which is heated locally by an external heat source, e.g. laser irradiation, is theoretically investigated in the absence of an applied magnetic field. 
We investigate the THz emission by solving the sine-Gordon, Maxwell, and thermal diffusion equation simultaneously and found that the emission power is largely enhanced by the local heating.
Based on the above analyses, we will clarify the optimum heating conditions for intense THz emission.

\begin{figure}[b]
\begin{center}
\includegraphics[width = 7.5cm]{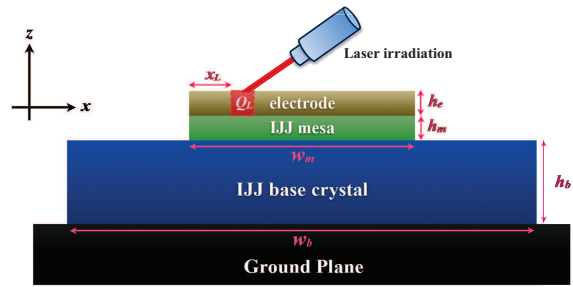}
\caption{ A two dimensional model of IJJ mesa which is locally heated by laser irradiation.
The IJJ mesa is fabricated on the IJJ base crystal. 
The mesa is covered by a metallic electrode.
The  laser  is irradiated on the upper electrode and locally increases the temperature of the mesa beneath the electrode. 
$x_L$ indicates the position of the heating spot. A DC voltage is applied to the IJJ mesa region.
$w_m = 60~\mu$m and  $w_b = 220~\mu$m are width of the IJJ mesa and the IJJ base crystal.  $h_e = 1~\mu$m, $h_m = 1~\mu$m and $h_b = 20~\mu$m 
are height of the electrode, the IJJ mesa and the IJJ base crystal. We take bath temperature $T_{bath} = 0.3~T_c$}
\label{f1}
\end{center}
\end{figure}

\section{Calculation method}
\ \  In Fig.~1, we show a  2D model of the IJJ emitter which is locally heated by laser irradiation.
The IJJ mesa is fabricated on the IJJ base crystal, and the mesa is covered by the upper electrode.
The upper electrode is locally heated by the focused laser beam, and the temperature of the IJJ mesa beneath the heating spot locally increases. 
In this calculation, we use the following dimensions : width of the electrode and the IJJ mesa $w_m = 60~\mu$m, width of the IJJ base crystal $w_b = 220~\mu$m, height of the electrode $h_e = 1~\mu$m,
height of the IJJ mesa $h_m = 1~\mu$m and height of the IJJ base crystal $h_b = 20~\mu$m\cite{Ozyuzer,jpsjKado,hot2Wang,TBenseman}.
The DC voltage is applied to the IJJ mesa region, and the base crystal is attached to the infinite ground plane.
%
In this study, we firstly solve the thermal diffusion equation with Joule heating in the IJJ mesa.
After that, we solve the sin-Gordon equation based on the obtained temperature distribution.

The thermal diffusion equation in IJJs is given by,
\begin{eqnarray}
0 = \frac{\partial}{\partial x} \Bigl[ \kappa_{ab}(T) \frac{\partial T}{\partial x} \Bigr] + \frac{\partial}{\partial z} \Bigl[ \kappa_c(T) \frac{\partial T}{\partial z}  \Bigr] 
  +  \frac{ j_{ex}^2}{\sigma_c(T)}.
\end{eqnarray}
Here, $T$ is the temperature, $\sigma_c$ is the $c$-axis conductivity in the IJJ mesa, and $\kappa_{ab}$, $\kappa_{c}$  is the thermal conductivity along the $ab$ plane and $c$ axis respectively. 
$j_{ex}$ is the external current density injected to the IJJ mesa region.
We assume the local heating by the external energy source and add the constant heat source $Q_{L}$ in the electrode region which is indicated by the red shaded area in Fig. 1.
The position of the heating spot is given by $x_{L}$, and we assume that the spot size is 5~$\mu$m.  In this study, we take $x_L$ = 13.3~$\mu$m.  
The dependence of emission power on the $x_L$ will be discussed in the future work.
%
We use the temperature dependent  $\kappa_{ab}$, $\kappa_c$ and $\sigma_c$ adopted in a previous theoretical study\cite{hotGross}.
In the thermal diffusion equation for upper electrode, we use an isotropic diffusion constant and take $\kappa = 20~$W/${\textrm m} \cdot {\textrm K}$. 
 We impose the boundary condition $T = T_{bath}$ at the boundary between IJJ base crystal and the ground plane.
 In this study, we take $T_{bath}= 0.3 T_c$.
The open boundary condition $\nabla T = 0$ is used  for other boundaries.

The dynamics of the phase differences in an IJJ mesa are described by the sine-Gordon equation within the in-phase approximation where all phase differences between IJJ layers are equal to the common phase difference $\phi$\cite{fdtdKoyama1, inphaseAsai},
\begin{eqnarray}
 \epsilon_c \frac{\partial^2 \phi}{\partial t^2}   =   c^2 \frac{\partial B_y}{\partial x} - \frac{1}{\epsilon_0 }  \left[  j_c (T) \sin{\phi} + \sigma_c (T) E_z -j_{ex} \right] ,
\end{eqnarray}
where,  $j_c(T)$ is the critical current density, $\epsilon_c$ is the dielectric constant of the junctions, $d$ is the thickness of insulating layers of IJJs, $c$ is the light velocity and $\Phi_0$ is the flux quantum.
The electromagnetic (EM) fields in the IJJs are given by $E_{z} (x,t) = \frac{\hbar}{2ed}  \frac{\partial \phi}{\partial t}$, $B_{y} (x,t) = \frac{\hbar}{2ed}\frac{\partial \phi}{\partial x}$.
We use the Ambegaokar-Baratoff relation for temperature dependence of the $j_c$\cite{reviewYTanaka}.
In this study, we take $\epsilon_c = 17.64$,  $d = 1.2$ nm and $j_c(0) = 4\cdot 10^2$ A/${\textrm {cm}}^2$\cite{jpsjKado,Inomata}. 


\begin{figure}[b]
\begin{center}
\includegraphics[width = 8cm]{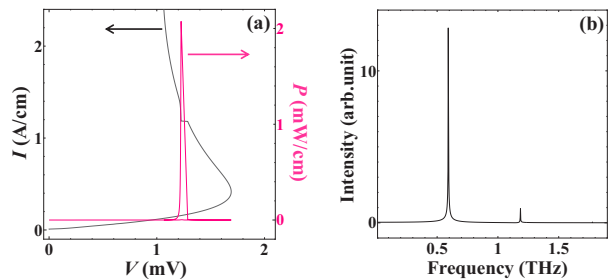}
\caption{Calculation results for for $Q_{L} = 0.3$ W/cm and $x_L$ = 13.3~$\mu$m. (a) The current $I$ and the emission power $P$ as a function of the voltage $V$.
(b) The frequency spectrum of the emission from the mesa.}
\label{f3}
\end{center}
\end{figure}

\section{Results and discussions}
\ \  We calculate the THz emission power by changing $Q_{L}$ for clarifying optimum ``heating power'' to achieve high power emission.
Figure 2 (a) shows the current $I$ and the emission power $P$ as a function of the voltage $V$ for $Q_{L}=0.3$ W/cm.   
The voltage here is divided by the number of IJJ layer in order to show the voltage applied to each IJJ layer.
%
%
At $V=1.23$ mV, the $I$ vs. $V$  curves show a small kink, and the $P$ vs. $V$ curve show a sharp peak.
Figure 2(b) shows the frequency spectrum of EM wave emitted by the mesa.
The peak frequency 0.59THz is equal to the AC Josephson frequency $f_J =  2eV/h$ and the cavity resonance frequency given by $f_c = cn/(2 \hspace{-0.8mm}\sqrt{\epsilon_c}w_m)$, where $n = 1$.
Moreover, we can  also see a small peak around 1.18THz which corresponds to the higher harmonics.
In this manner, strong THz wave is emitted from the mesa when the cavity resonant EM mode appears in the mesa.

\begin{figure}
\begin{center}
\includegraphics[width = 8cm]{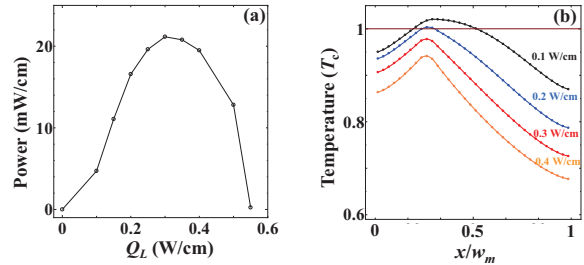}
\caption{(a) The emission power as a function of $Q_{L}$ for $x_{L} = 13.3$ $\mu$m. (b)The distribution of $T$ in the mesa for $Q_{L} = 0.1, 0.2, 0.3, 0.4$ W/cm.}
\label{f1}
\end{center}
\end{figure}

In Fig 3(a), we shows the emission power as a function of $Q_{L}$ to discuss the optimum heating power for high power emission.
This figure shows a peak around $Q_{L} = 0.3$ W/cm.
The spatial distribution of the temperature $T$ in the mesa during the intense emission are shown in Fig. 3(b).
The hot spot where $T$  is locally higher than $T_c$ appears for $Q_{L} \leq 0.2$ W/cm.
Remarkably the emission power $P$ is largely enhanced by external heat source $Q_L$ in comparison to the case without the external local heating as shown in Fig 3 (a).

Our results indicate external local heating is powerful method to achieve high power THz emission. 
Of particular note is that the strongest emission is obtained around $Q_L =0.3$ W/cm.
In this case,  the temperature of the heating spot is slightly lower than $T_c$, and thus the hot spot ($T>T_c$) is not formed.
The change of $j_c$ becomes remarkable when $T$ is slightly below $T_c$ as expected from the temperature dependence of $j_c$\cite{reviewYTanaka}.
Therefore the drastic $j_c$ modulation by the local heating strongly excites the THz Josephson plasma wave inside the IJJ mesa. 
On the other hand, for $Q_L < 0.3$~W/cm, a hot spot region is formed during the emission. Since a hot spot region does not contribute to THz emission because the $j_c = 0$  in this region,
the formation of hot spot results in the reduction of the emission power.
Conversely, for  $Q_L > 0.3$~W/cm, the mesa temperature becomes much lower than $T_c$.  In this case 
the modulation of $j_c$ becomes small because the $j_c$ has little temperature dependence unless $T$ is close to $T_c$.
Consequently, the excitation of the THz Josephson plasma wave becomes weak.
Thus, we can conclude that {\it the local heating which keeps the mesa temperature ``slightly lower'' than $T_c$ is preferable for high power emission}.
Moreover, we would like to emphasize that our result is consistent with the recent experimental study which reported the strongest emission is realized when $T$ is slightly lower than $T_c$\cite{TBenseman}. 
In addition, we have compared the temperature distributions in our calculation with those in the experiments, and we found that our results are qualitatively similar to those reported in Ref.~\cite{TBenseman}.


\section{Conclusion}

\ \  We have investigated THz emission from the IJJ mesa  which is locally heated by an external heat source. 
We clarified the optimum heating condition in order to achieve high power THz emission. 
The key point to design an intense THz emitter is  a control of the heating power to keep mesa temperature ``slightly lower'' than $T_c$.
We believe that the precise control of local temperature based on our guidelines will be a one of pragmatic way to realize the practical THz emitter using intrinsic Josephson junctions.

We wish to thank K. Kadowaki, I. Kakeya, T. Kashiwagi, F. Nori, H. Minami Y. Ota, M. Tachiki, M. Tsujimoto and H. B. Wang for fruitful discussions and comments.

\


%



\end{document}